\documentclass[%
 reprint,
 superscriptaddress,
 amsmath,amssymb,
 aps,
 prc
]{revtex4-2}

\usepackage[utf8]{inputenc}
\usepackage{graphicx}
\usepackage{dcolumn}
\usepackage{bm}
\usepackage{hyperref}
\usepackage{amsfonts}
\usepackage{mathrsfs}
\usepackage{bm}
\usepackage{rotating} 
\usepackage{MnSymbol}
\usepackage{color}
\usepackage[dvipsnames]{xcolor}
\usepackage[active]{srcltx}
\usepackage{multirow}
\usepackage{caption} 
\usepackage{comment} 
\usepackage{physics} 
\usepackage{soul}
\newcommand{\si}[1]{\,\rm{#1}}
\newcommand{\sn}[1]{\times10^{#1}}
\newcommand{\nn}[2]{^{#1}\rm{#2}}
\newcommand{\nhl}[2]{^{#1}_{\Lambda}\rm{#2}}
\newcommand{\spl}[3]{{#1}{#2}_{{#3}/2}}
\begin{document}

\date{\today}

\title{Gamow shell model description of the hypernuclei}


\author{Alan Cruz Dassie}
\affiliation{Grand Acc\'{e}l\'{e}rateur National d'Ions Lourds (GANIL), CEA/DSM - CNRS/IN2P3, BP 55027, F-14076 Caen Cedex, France}

\author{Emiko Hiyama}
\affiliation{Department of Physics, Tohoku University, Aoba 6-3, Sendai, 980-8578, Miyagi, Japan}
\affiliation{Nishina Center, RIKEN, Hirosawa 2-1, Wako, 351-0198, Saitama, Japan}
\affiliation{Universit\'{e} Paris-Saclay, CNRS/IN2P3, IJClab, 91405 Orsay, France}

\author{Nicolas Michel}
\affiliation{CAS Key Laboratory of High Precision Nuclear Spectroscopy,Institute of Modern Physics, Chinese Academy of Sciences, Lanzhou 730000, China}
\affiliation{School of Nuclear Science and Technology, University of Chinese Academy of Sciences, Beijing 100049, China}

\author{Marek P{\l}oszajczak}
\affiliation{Grand Acc\'{e}l\'{e}rateur National d'Ions Lourds (GANIL), CEA/DSM - CNRS/IN2P3, BP 55027, F-14076 Caen Cedex, France}



\begin{abstract}
\begin{description} 
  \item[Background] Hypernuclear physics studies baryon interactions and the structure of exotic atomic nuclei, where strangeness plays a key role in dense matter. High-resolution $\gamma$-ray spectroscopy (e.g., Hyperball at BNL/KEK) and upcoming facilities (J-PARC, JLab) have provided precise data on $p$-shell hypernuclei, constraining $YN$ potentials.
  \item[Purpose] We applied the Gamow shell model to $p$-shell hypernuclei to systematically investigate the role of the $\Lambda N$ interaction and its components, and how it affects the mean field of nucleons.
  \item[Method] The Gamow shell model extends the standard shell model by replacing the harmonic oscillator basis with the Berggren ensemble, treating bound, resonant, and continuum states on equal footing. The effective Hamiltonian includes Woods-Saxon core potentials and two-body interactions (central, spin-orbit, tensor) optimized to experimental data.
  \item[Results] Our calculations reproduce and predict binding energies, excitation spectra, and densities for hypernuclei from ${}^{5}_{\Lambda}\mathrm{He}$ to ${}^{16}_{\Lambda}\mathrm{O}$ using the same $\Lambda N$ interaction. The tensor force significantly impacts excited-state splittings in bound systems, while its effect is suppressed in unbound cases due to the continuum coupling.
  \item[Conclusions] The Gamow shell model provides a unified framework for hypernuclei, capturing the interplay between bound, resonant, and continuum states. This work lays the foundation for extending the model to heavier and multi-strange systems.
\end{description}
\end{abstract}

\maketitle

\section{Introduction}

Hypernuclear physics stands at the frontier of nuclear and particle physics, offering a unique window into the study of baryon-baryon interactions, the structure of exotic nuclei, and the role of strangeness in dense matter. Hypernuclei ---nuclear systems containing one or more hyperons (e.g., $\Lambda$, $\Sigma$, $\Xi$)--- provide an opportunity to probe the hyperon-nucleon ($YN$) and hyperon-hyperon ($YY$) forces, which are critical for understanding the equation of state (EoS) of neutron stars and the behavior of matter under extreme conditions \cite{galRev.Mod.Phys.2016,vidanaProc.R.Soc.Math.Phys.Eng.Sci.2018,chenChinesePhys.Lett.2025}. Recent years have witnessed remarkable progress in both experimental and theoretical hypernuclear physics. On the experimental front, high-resolution $\gamma$-ray spectroscopy \cite{tamuraNuclearPhysicsA2013},
emulsion experiment \cite{hiyamaAnnu.Rev.Nucl.Part.Sci.2018} and femtoscopic experiments
\cite{STAR,ALICE,ALICE1} have provided many epoch-making data related to $YN$ and $YY$ interactions. Theoretically, using these experimental data, information on spin-dependent terms of $YN$ interaction and spin-independent term of $YY$ interactions have been extracted within shell model~\cite{millenerNuclearPhysicsA2005,millenerNuclearPhysicsA2008,millenerNuclearPhysicsA2013} and cluster calculation \cite{hiyamaPhys.Rev.C2009}.
Currently, it is important to extract information on $\Lambda N-\Sigma N$ coupling term which is related to effective three-body force and charge symmetry breaking effect in $\Lambda$ hypernuclei.
For this study, one of neutron-rich $\Lambda$
hypernuclei, $^7_{\Lambda}$He has been observed at
Jlab \cite{HKSJLabE01-011:2012sgn}
and the structure calculation determined the $\Lambda N-\Sigma N$ coupling in this nucleus \cite{hiyamaPhys.Rev.C2015}.
However, for the study of neutron star, it is
crucial to study structure of heavier $\Lambda$
hypernuclei.
For this purpose,
it is planned to observe energy spectra of $^{40,48}_{\Lambda}$K using $^{40,48}$Ca targets by
$(e,e'K^+)$reaction \cite{JLabProposal-E12-24-013}.
Under this situation, it is highly requested to
predict the energy spectra of these $\Lambda$ hypernuclei with the reliable $\Lambda N$ interaction which reproduce energy spectra of observed $p$ to $sd$-shell $\Lambda$ hypernuclei.
Such a calculation in a relatively small model space was reported in Refs. \cite{millenerNuclearPhysicsA2005,millenerNuclearPhysicsA2008,millenerNuclearPhysicsA2013}.

To predict energy spectra in heavier systems,
it is highly desirable to perform shell model calculation in larger configuration space.
One approach that can meet such requirements is,
 the Gamow shell model (GSM) which has emerged as a powerful framework for describing open quantum systems, such as nuclei and hypernuclei near or beyond particle emission thresholds \cite{michelPhys.Rev.Lett.2002,idbetanPhys.Rev.Lett.2002,michelJ.Phys.GNucl.Part.Phys.2009,jaganathenJ.Phys.:Conf.Ser.2012,michel2021}. By replacing the traditional harmonic oscillator basis with the Berggren ensemble, GSM enables a unified treatment of all quantum states within the complex momentum plane. This approach is particularly well-suited for hypernuclei, where the $\Lambda$ hyperon, unconstrained by the Pauli exclusion principle relative to nucleons, can occupy deeply bound orbitals and exhibit unique structural features, such as halo or skin configurations \cite{myoPhys.Rev.C2023,knollPhysicsLettersB2026}.

The description of hypernuclei in an extended single-particle basis including bound, resonance, and complex-energy scattering states of the Berggren ensemble requires a development of the appropriate $YN$ and $NN$ effective interaction. It is the primary objective of this work to construct such an interaction based on the available experimental information on spectra and binding energies in light nuclei and hypernuclei.
The effective Hamiltonian of hypernuclei includes a Woods-Saxon core potential and a two-body interaction modeled as a sum of central, spin-orbit, and tensor components, with parameters optimized to reproduce experimental data for selected nuclei and hypernuclei.

We shall present a systematic application of the GSM to the description of isotopic chains in $p$-shell nuclei and hypernuclei, focusing on the role of a specific $\Lambda N$ interaction. Our approach builds upon the GSM’s ability to handle both bound and unbound states what is essential in long chains of isotopes. Based on the available data, we will investigate the $A$-dependence of the $\Lambda N$ interaction to determine the evolution of its parameters from the beginning of the  $p$-shell to the $sd$-shell.

This paper is organized as follows. In Section \ref{sec:theory}, we outline the theoretical framework of the GSM and the construction of the effective Hamiltonian for hypernuclei. In Section \ref{sec:nucleon-sector}, we present our results of the GSM description of nuclei where we determined the $NN$ interaction. In Section \ref{sec:n-lambda_interaction} we determine the effective $\Lambda$-core and $\Lambda N$ interactions for GSM calculation
to reproduce spectra and binding energies of He, Li and Be $\Lambda$-hypernuclei.
In the next Section \ref{sec:hypernuclei-spectra}, we check validity of the so determined $\Lambda N$ interaction in selected heavier $p$-shell $\Lambda$ hypernuclei such as $^{12}_{\Lambda}$B, $^{12}_{\Lambda}$C and $^{16}_{\Lambda}$O, and in the Section \ref{sec:hypernuclei-densities}, we 
show accuracy of the calculated GSM wavefunctions for radial densities and root-mean-square radii of selected hypernuclei. Finally, Section \ref{sec:conclusions} summarizes our conclusions and outlines future directions for extending the GSM approach to heavier hypernuclear systems.

\section{Theoretical framework} \label{sec:theory}

\subsection{Gamow shell model: the general method}

The GSM is a configuration interaction framework formulated to describe the properties of open quantum systems, such as nuclei and hypernuclei near or beyond particle emission thresholds \cite{michel2021,liPhysicsLettersB2025}. As a direct generalization of the standard nuclear shell model, it replaces the conventional harmonic oscillator basis with the Berggren ensemble \cite{berggrenNucl.Phys.A1968}, which encompasses single-particle bound states, resonances, and the non-resonant scattering continuum. This approach allows for the unified treatment of all categories of quantum states on an equal footing within the complex momentum plane. The GSM provides a description of the baryon system as an isolated entity that preserves unitarity across the entire range of binding energies.

The necessity for such an extension arises from the limitations of traditional shell model approaches when applied to weakly bound or unbound systems. In these cases, the continuum can significantly renormalize the effective interaction and modify observables like binding energies and transition strengths, even for bound states. For unbound states, such as resonances above particle emission thresholds, the wave functions explicitly contain components of the scattering continuum, which cannot be neglected. Near particle emission thresholds, the interplay between bound and continuum states leads to phenomena such as threshold cusps, halo structures, and enhanced collectivity, none of which can be captured by models confined to bound-state bases \cite{michelJ.Phys.GNucl.Part.Phys.2009,okolowiczPhysicsReports2003,Okolowicz2013}.

The mathematical foundation of the GSM is a departure from the standard Hilbert space, which cannot accommodate state vectors that exhibit exponential growth or decay, such as resonances. Instead, the model is formulated within the Rigged Hilbert Space (RHS) \cite{gelfand1964generalized,Maurin1968,10.1063/1.4758925}. This framework allows for the inclusion of Gamow states —generalized eigenvectors of the Hamiltonian with complex eigenvalues— that are not square-integrable and thus do not belong to the standard Hilbert space.

Gamow states are associated with the poles of the $S$-matrix and exhibit exponentially increasing asymptotics in coordinate space. For a resonance with complex momentum $k_n = \gamma_n - i\kappa_n$ (where $\gamma_n, \kappa_n > 0$), the radial wave function behaves as:
\begin{equation}
u_{n\ell j}(r) \sim C e^{i k_n r} = C e^{i \gamma_n r} e^{\kappa_n r}, \quad r \to \infty.
\end{equation}
This divergence is compensated by the exponential decay in time of the resonance, $e^{-i E_n t/\hbar} = e^{-i (E_0 - i\Gamma/2) t/\hbar} = e^{-i E_0 t/\hbar} e^{-\Gamma t/(2\hbar)}$, ensuring that the probability density remains well-behaved for physical observables. The normalization of Gamow states is defined using the external complex scaling method, which ensures that matrix elements involving resonant states are finite and well-defined.

The completeness relation on the Berggren ensemble is expressed as a sum over discrete bound states and narrow resonances, plus a contour integral over a background of non-resonant scattering states in the complex momentum $k$-plane \cite{berggrenNucl.Phys.A1968}:
\begin{align}
    & \sum_{n=b,d} \ketbra{\Tilde{u}_n}{u_n} + \int_{\mathfrak{L}^+} \braket{\Tilde{u}(k)}{u(k)} dk = 1
    \label{eq:berggren_completeness}
\end{align}
where $b$ and $d$ denotes the bound states and selected decaying resonant states, respectively. $\mathfrak{L}^+$ is the complex contour of scattering states, which must include the resonance states present in the discrete sum. In practice, the contour is truncated at a $k_\mathrm{max}$ value, typically  $2-3\si{fm^{-1}}$, and discretized, so that only a finite number of single particle states are present in numerical calculations \cite{jaganathenPhys.Rev.C2017}.

To describe the many-body system, the GSM constructs Slater determinants from these single-particle Berggren states \cite{jaganathenPhys.Rev.C2017}. The GSM utilizes Cluster Orbital shell model (COSM) coordinates, where valence particle coordinates are defined relative to the center of mass of an inert core. The resulting many-body Hamiltonian includes a recoil term that accounts for the core's motion relative to the valence baryons \cite{jaganathenJ.Phys.:Conf.Ser.2012}. 

The GSM Hamiltonian reads with COSM coordinates \cite{jaganathenPhys.Rev.C2017,michel2021}
\begin{align}
    \hat{H}_\mathrm{GSM} & = \sum_{i=1}^{A_\mathrm{val}} \left( \frac{p_i^2}{2\mu_i} + \hat{U}^{(c)}_i \right) + \sum_{i<j}^{A_\mathrm{val}} \left( \hat{V}_{ij}^{(\mathrm{res})} + \frac{p_i\cdot p_j}{M_c} \right),
    \label{eq:gsm_hamiltonian}
\end{align} 
where $\mu_i$, $M_c$ stand for the effective mass of the baryon and the mass of core, respectively. $\hat{U}^{(c)}_i$ is the single-particle core potential acting on the i-th particle, and $\hat{V}_{ij}^{(\mathrm{res})}$ is the interaction between valence nucleon-nucleon and nucleon-hyperon pairs.

The hermitian GSM Hamiltonian takes the form of a complex-symmetric matrix to diagonalize in the Berggren basis due to the outgoing boundary conditions of the Gamow states. However, its eigenvalues correspond to the physical energies and widths of the many-body states. A key challenge in GSM is to distinguish physical resonances from the non-resonant continuum in the spectrum of the Hamiltonian. Physical states are unambiguously identified using an overlap method \cite{michelJ.Phys.GNucl.Part.Phys.2005}. Physical resonances are also stable with respect to small deformations of the scattering contour $\mathfrak{L}^+$, whereas non-resonant states ``walk" in the complex energy plane as the contour changes.

In the context of hypernuclei, the GSM treats $\Lambda$-hyperons and nucleons on the same footing within the configuration interaction framework \cite{liPhysicsLettersB2025}. Unlike nucleons, $\Lambda$-hyperons are not subject to the Pauli exclusion principle relative to nucleons, allowing them to occupy the deeply bound $\spl{1}{s}{1}$ orbital.

\subsection{GSM effective hamiltonian} \label{sec:gsmeffective}

The GSM effective Hamiltonian for hypernuclei includes a Woods-Saxon potential for the baryon-core interaction, with central, spin-orbit, and Coulomb (for protons) components:
\begin{align}
    U_c(r) & = V_\text{WS} f(r) - 4V_{\ell s} \frac{1}{r} \frac{df(r)}{dr} (\vb*{\ell}\cdot\vb*{s}) + U_\mathrm{Coul}(r)
    \label{eq::onebody_interaction}
\end{align}
where $f(r) = -(1+\exp[(r-R_0)/a])^{-1}$ is the standard WS form factor, $V_\text{WS}$ is the central potential depth, $R_0$ is the WS radius, $V_{\ell s}$ is the spin-orbit strength, and $a$ is the WS potential diffuseness. The latter parameters are applied to all valence protons, neutrons, and $\Lambda$-hyperons. 

For different cases, the strengths $V_0$ for neutrons and protons might need to be smoothly modified from isotope to isotope according to the following isospin dependent value $V_\text{WS}(\tau) = \Bar{v}_\text{WS}^\tau (1\pm\chi(N-Z)/A$ with $\tau=\{p,n\}$, $+$ for protons and $-$ for neutrons. The values of $\Bar{v}_0^\tau$ and $\chi$ will be determined for each isotopic chain. The Coulomb potential is generated by a spherical Gaussian charge distribution for the frozen core.

The baryon-baryon interaction is modeled as a sum of the components central ($V_C$), spin–orbit ($V_{LS}$) and tensor ($V_T$) for the nucleon-nucleon case only \cite{jaganathenPhys.Rev.C2017,hiyamaPhys.Rev.C2006}:
\begin{align}
    V = V_C + V_{LS} + V_{T}
\end{align}
in a way that all kind of baryons interact by the same type of interaction.

For the nucleon-nucleon ($NN$) interaction, the central, spin-orbit and tensor terms follow the Furutani-Horiuchi–Tamagaki (FHT) form \cite{furutaniProgressofTheoreticalPhysics1979} rewritten in terms of the spin-isospin projectors $\Pi_{ST}$ \cite{ring2004,jaganathenPhys.Rev.C2017}. The $\Lambda N$ terms also follow the FHT form as stated by the implementation from Ref. \cite{hiyamaPhys.Rev.C2006} with the inclusion of the anti-symmetric spin-orbit term (ALS):
\begin{align}
    V^{\Lambda N}_C(r) & = \sum_\alpha \bar{v}^{\Lambda N}_\alpha \sum_{i=1}^3 v^C_{\alpha,i} e^{-(r/\beta^C_i)^2} \Pi_\alpha \label{eq.central_lambda-n}\\
    V^{\Lambda N}_{LS}(r) & = \bar{v}^{LS}_\mathrm{to} \sum_{i=1}^2 v^{LS}_{\mathrm{to},i} \left[ \vb{L}\cdot(\vb{s}_\Lambda + \vb{s}_N) + \right. \label{eq.spinorbit_lambda-n}\\
    &\,\,\,\,\,\,\,\,\,\,\,\,\,\,\,\,\,\,\,\,\,\,\,\,\,\,\,\, \left. + \eta_{ALS} \vb{L}\cdot(\vb{s}_\Lambda - \vb{s}_N) \right] e^{-(r/\beta^{LS}_i)^2} \Pi_\mathrm{to}  \nonumber\\
    V^{\Lambda N}_T(r) & = S_{\Lambda N} \sum_{i=1}^3 \left[ \bar{v}^T_\mathrm{te} v_{\mathrm{te},i}^T e^{-(r/\beta^T_i)^2} \Pi_\mathrm{te} + \right. \label{eq.tensor_lambda-n} \\
    &\,\,\,\,\,\,\,\,\,\,\,\,\,\,\,\,\,\,\,\,\,\,\,\,\,\,\,\, + \left. \bar{v}^T_\mathrm{to} v_{\mathrm{to},i}^T e^{-(r/\beta^T_i)^2} \Pi_\mathrm{to} \right] \nonumber
\end{align}
where $r\equiv r_{\Lambda N_j}$ is the distance between the $\Lambda$ and a valence nucleon, $\alpha = \{\mathrm{se,to,so,te}\}$ and $S_{\Lambda N}=3\left(\vb{s}_\Lambda\cdot\hat{r}\right)\left(\vb{s}_N\cdot\hat{r}\right) - \vb{s}_\Lambda \cdot\vb{s}_N$. The even(odd) spin-dependent part of the central interaction is $(\bar{v}^{\Lambda N}_\mathrm{te(to)} v^C_{\mathrm{te(to)},i}-\bar{v}^{\Lambda N}_\mathrm{se(so)} v^C_{\mathrm{se(so)},i})/4$, while the even(odd) spin-independent part is $(3\bar{v}^{\Lambda N}_\mathrm{te(to)} v^C_{\mathrm{te(to)},i}+\bar{v}^{\Lambda N}_\mathrm{se(so)} v^C_{\mathrm{se(so)},i})/4$. The strengths $\bar{v}^{\Lambda N/LS/T}$ of each component and the antisymmetric spin-orbit factor $\eta_{ALS}$ will be optimized. The parameters $v_{\alpha,i}$ and $\beta_i$ for central and spin-orbit parts are taking from Ref. \cite{hiyamaPhys.Rev.C2009} and from Ref. \cite{furutaniProgressofTheoreticalPhysics1979} for tensor part.
The central and spin-orbit strengths were adjusted to simulate the characteristics of the Nijmegen model NSC97f \cite{PhysRevC.59.21} with the $\Lambda N-\Sigma N$ coupling effects renormalized into $\Lambda N-\Lambda N$ parts for the $\nhl{7}{Li}$ hypernuclei. 

\section{Applications} \label{sec:aplication}

\subsection{Determination of the nucleon-nucleon interaction} \label{sec:nucleon-sector}

To describe hypernuclei accurately, we first determine the nucleon-nucleon interaction by fitting the Hamiltonian parameters to experimental data for selected nuclei. We consider here the three chain of isotopes (He, Li, Be), the mirror nuclei with mass $A=11$ and the $\nn{15}{O}$.

For the three chains of isotopes, the model space in the nucleon sector consists of the single-particle states $\spl{1}{p}{3}$ and $\spl{1}{p}{1}$ with particle-hole excitations up to $8\hbar\omega$, and a maximum of two valence particles in the continuum. For the systems $\nn{10,11}{B}$, $\nn{11,12}{C}$, and $\nn{15}{O}$, the single-particle states $\spl{1}{d}{5}$ and $\spl{2}{s}{1}$ are added to the model space to allow for multi-shell configurations of the heaviest hypernuclei. The optimized core-nucleon interaction strengths and their statistical uncertainties \cite{Dobaczewski_2014} of all the sets of nuclei are shown in Table \ref{tab:sec:nucleonsector:ob_optimized}. For the three chains of isotopes, as mentioned in Sec. \ref{sec:gsmeffective}, the WS strengths in Table \ref{tab:sec:nucleonsector:ob_optimized} follow an isospin dependence given by:
\begin{align}
    V_\text{WS}(\tau) & = \Bar{v}_\text{WS}^\tau \left[ 1 + \chi \left( \frac{N - Z}{A} \right) \right]
    \label{eq:ws_isospin_dependence}
\end{align}
with $\chi_\mathrm{He/Be} = 0.045$ and $\chi_\mathrm{Li} = 0.04$.

\begin{table}[h!tb]
	\centering
    \setlength{\tabcolsep}{0.5em}
    \renewcommand{\arraystretch}{1.5} 
	\caption{One-body WS and $\ell s$ optimized strengths and their statistical uncertainties (in MeV). The diffusivity is $a=0.65\si{fm}$ and the radius is fixed to $R_0 = 1.25A^{1/3}\si{fm} \simeq 2\si{fm}$ where A=4 is related to the $\nn{4}{He}$ core. Some uncertainties are written with a decimal point to denote their full value, e.g., $49.06(6.)$ means $49.06 \pm 6.00$.}
	\label{tab:sec:nucleonsector:ob_optimized}
	\begin{tabular}{c|c|c|c|c}
        \hline\hline
        $V[CN]$ & $V_\text{WS}^p$ & $V_\text{WS}^n$ & $V_\text{so}^p$ & $V_\text{so}^n$ \\\hline
        He-chain & $-$ & $49.06(6.)$ & $-$ & $5.56(3.)$ \\\hline
        Li-chain & $48.18(2.)$ & $56.27(4.)$ & $2.06(1.)$ & $4.35(2.)$ \\\hline
        Be-chain & $51.95(7.)$ & $48.86(7.)$ & $1.97(1.)$ & $4.54(2.)$ \\\hline
        A=10-12 & $53.12(2.)$ & $53.41(2.)$ & $5.17(5.)$ & $4.64(52)$ \\\hline
        $\nn{15}{O}$ & $51.12(2.)$ & $51.41(2.)$ & $5.07(4.)$ & $4.54(4.)$ \\
        \hline\hline
	\end{tabular}
\end{table}

\begin{table}[h!tb]
    \centering
    \renewcommand{\arraystretch}{1.5} 
    \setlength{\tabcolsep}{0.3em}
	\caption{Optimized strengths and their statistical uncertainties (in MeV) of the two-body central (C), spin-orbit (SO) and tensor (T) parts that multiplies the FHT interaction parameters from Ref. \cite{furutaniProgressofTheoreticalPhysics1979} on the spin-isospin projector form. The triplet-even spin-orbit and tensor potentials are fixed to zero. Some uncertainties are written with a decimal point to denote their full value, e.g., $-7.83(7.)$ means $-7.83 \pm 7.00$.}
	\label{tab:sec:nucleonsector:tb_optimized}
	\begin{tabular}{c|c|c|c|c}
        \hline\hline
		$V_\alpha[NN]$ & $V^\mathrm{C}_\text{to}$ & $V^\mathrm{C}_\text{te}$ & $V^\mathrm{C}_\text{so}$ & $V^\mathrm{C}_\text{se}$\\ \hline
		He-chain & $0$ & $-$ & $-$ & $-7.83(7.)$  \\ \hline
        Li-chain & $-5.93(81.)$ & $-6.94(11.)$ & $-37.48(64.)$ & $-4.44(6.)$ \\ \hline
        Be-chain & $-29.15(32.)$ & $-8.02(34.)$ & $-46.74(67.)$ & $-6.29(22.)$ \\ \hline
        A=10-12 & $10.56(15.)$ & $-6.78(8.)$ & $-6.19(5.)$ & $-2.46(5.)$ \\ \hline
        $\nn{15,16}{O}$ & $11.01(46.)$ & $-11.05(18.)$ & $-6.33(14.)$ & $4.90(9.)$  \\ \hline\hline
        $V_\alpha[NN]$ & \multicolumn{2}{c|}{$V^\mathrm{SO}_\text{to}$} & \multicolumn{2}{c}{$V^\mathrm{T}_\text{to}$}\\ \hline
        He-chain        & \multicolumn{2}{c|}{$-32.53(85.)$} & \multicolumn{2}{c}{$-11.16(5.)$} \\\hline
        Li-chain        & \multicolumn{2}{c|}{$-2459(5950)$} & \multicolumn{2}{c}{$17.59(85.)$} \\\hline
        Be-chain        & \multicolumn{2}{c|}{$1703(2530)$} & \multicolumn{2}{c}{$21.83(38.)$} \\\hline
        A=10-12      & \multicolumn{2}{c|}{$960(996)$} & \multicolumn{2}{c}{$5.12(7.)$} \\\hline
        $\nn{15,16}{O}$ & \multicolumn{2}{c|}{$794(1084)$} & \multicolumn{2}{c}{$-26.46(13.)$} \\
        \hline\hline
	\end{tabular}
\end{table}

\begin{table}[h!tb]
    \centering
    \renewcommand{\arraystretch}{1.32} 
    \caption{Calculated and experimental \cite{nationalnucleardatacenter} energies (in MeV) with respect to the $\nn{4}{He}$ core and widths (in MeV) of all the optimized states of the selected nuclei. Experimental uncertainties are given in parentheses.}
    \begin{tabular}{cc|@{\hspace{0.5em}}c@{\hspace{0.5em}} @{\hspace{0.5em}}c@{\hspace{0.5em}}|@{\hspace{0.3em}}c@{\hspace{0.3em}}c}
        \hline\hline
        Nucleus & State & $E_\mathrm{GSM}$ & $E_\mathrm{Exp}$ & $\Gamma_\mathrm{GSM}$ & $\Gamma_\mathrm{Exp}$ \\\hline
        $\nn{5}{He}$ & $3/2^-$ & $0.712$ & $0.736$ & $0.594$ & $0.648$ \\
		$\nn{5}{He}$ & $1/2^-$ & $1.905$ & $2.006$ & $3.463$ & $5.570$ \\
		$\nn{6}{He}$ & $0^+$ & $-1.010$ & $-0.975$ & $-$ & $-$ \\
		$\nn{6}{He}$ & $2^+$ & $0.824$ & $0.822$ & $0.089$ & $0.113(20)$ \\
		$\nn{7}{He}$ & $3/2^-$ & $-0.528$ & $-0.566$ & $0.121$ & $0.150(20)$  \\
		$\nn{7}{He}$ & $5/2^-$ & $1.685$ & $2.354$ & $0.904$ & $1.990(170)$  \\
		$\nn{8}{He}$ & $0^+$ & $-3.532$ & $-3.101$ & $-$ & $-$ \\
		$\nn{8}{He}$ & $2^+$ & $-0.791$ & $-0.001$ & $-$ & $-$ \\ \hline
        $\nn{5}{Li}$ & $3/2^-$ & $1.972$ & $1.966$ & $2.039$ & $1.060$ \\
		$\nn{5}{Li}$ & $1/2^-$ & $2.453$ & $3.456$ & $3.424$ & $3.780$ \\
		$\nn{6}{Li}$ & $1^+$ & $-3.549$ & $-3.698$ & $-$ & $-$ \\
		$\nn{6}{Li}$ & $3^+$ & $-1.629$ & $-1.512$ & $-$ & $-$ \\
		$\nn{6}{Li}$ & $0^+$ & $0.013$ & $-0.135$ & $0.031$ & $-$ \\
		$\nn{7}{Li}$ & $3/2^-$ & $-10.726$ & $-10.949$ & $-$ & $-$  \\
		$\nn{7}{Li}$ & $1/2^-$ & $-10.257$ & $-10.472$ & $-$ & $-$  \\
		$\nn{8}{Li}$ & $2^+$ & $-13.591$ & $-12.982$ & $-$ & $-$ \\
		$\nn{8}{Li}$ & $1^+$ & $-12.770$ & $-12.001$ & $-$ & $-$ \\ 
		$\nn{9}{Li}$ & $3/2^-$ & $-16.605$ & $-17.044$ & $-$ & $-$ \\
		$\nn{9}{Li}$ & $1/2^-$ & $-15.494$ & $-14.353$ & $-$ & $-$ \\\hline
        $\nn{6}{Be}$ & $0^+$ & $1.593$ & $1.372$ & $0.040$ & $0.092(6)$ \\
		$\nn{6}{Be}$ & $2^+$ & $2.725$ & $3.042$ & $0.836$ & $1.160(60)$ \\
		$\nn{7}{Be}$ & $3/2^-$ & $-9.540$ & $-9.305$ & $-$ & $-$ \\
		$\nn{7}{Be}$ & $1/2^-$ & $-8.565$ & $-8.876$ & $-$ & $-$ \\
		$\nn{8}{Be}$ & $0^+$ & $-28.152$ & $-28.204$ & $-$ & $-$  \\
		$\nn{8}{Be}$ & $2^+$ & $-25.169$ & $-25.174$ & $-$ & $-$  \\
		$\nn{9}{Be}$ & $3/2^-$ & $-29.526$ & $-29.868$ & $-$ & $-$ \\
		$\nn{9}{Be}$ & $5/2^-$ & $-27.258$ & $-27.439$ & $0.001$ & $0.001$ \\
		$\nn{9}{Be}$ & $1/2^-$ & $-27.203$ & $-27.088$ & $0.001$ & $1.100(12)$ \\ \hline
        $\nn{10}{B}$ & $3^+$ & $-36.447$ & $-36.455$ & $-$ & $-$\\
        $\nn{10}{B}$ & $1^+$ & $-35.747$ & $-35.737$ & $-$ & $-$\\
        $\nn{11}{B}$ & $3/2^-$ & $-47.896$ & $-47.909$ & $-$ & $-$\\
        $\nn{11}{B}$ & $1/2^-$ & $-45.805$ & $-45.785$ & $-$ & $-$\\
        $\nn{11}{B}$ & $5/2^-$ & $-43.785$ & $-43.464$ & $-$ & $-$\\
        $\nn{11}{C}$ & $3/2^-$ & $-45.170$ & $-45.145$ & $-$ & $-$\\
        $\nn{11}{C}$ & $1/2^-$ & $-43.128$ & $-43.145$ & $-$ & $-$\\
        $\nn{11}{C}$ & $5/2^-$ & $-41.191$ & $-40.827$ & $-$ & $-$\\
        $\nn{12}{C}$ & $0^+$ & $-63.825$ & $-63.866$ & $-$ & $-$\\ \hline
        $\nn{15}{O}$ & $1/2^-$ & $-83.653$ & $-83.660$ & $-$ & $-$\\
        $\nn{15}{O}$ & $1/2^+$ & $-78.475$ & $-78.477$ & $-$ & $-$\\
        $\nn{15}{O}$ & $3/2^-$ & $-77.492$ & $-77.483$ & $-$ & $-$\\
        $\nn{15}{O}$ & $5/2^-$ & $-74.167$ & $-74.172$ & $-$ & $-$\\
        $\nn{15}{O}$ & $7/2^-$ & $-74.002$ & $-73.998$ & $-$ & $-$\\
        \hline\hline
    \end{tabular}
    \label{tab:sec:nucleonic-part:optimizedenergies-widths}
\end{table}

Since the particle emission threshold for certain isotopes of the He-, Li-, and Be-chains are close to the ground state, and some isotopes are unbound, the Berggren basis is used for these chains. This basis includes the $\spl{1}{p}{3}$ and $\spl{1}{p}{1}$ poles and their associated scattering continua, as well as real scattering $\ell=0$ states for both protons and neutrons, and a real scattering continuum for the $\spl{1}{d}{5}$ partial wave for neutrons to account for possible excited states of the heavier isotopes. When the low-lying many-body states are well bound with respect to neutron, proton, or $\Lambda$-hyperon emission, the basis potential that generates the Berggren basis is modified to ensure all poles are bound, using contours along the real axis of the momentum plane discretized by $30$ mesh points up to $k_\mathrm{max}=2\si{fm^{-1}}$. This applies to all lithium isotopes, except for $\nn{6}{Li}$. Particular consideration must be given when the many-body states are unbound or loosely bound, such as the helium isotopes. In these cases, one or both $\ell=1$ poles are moved to the fourth quadrant of the complex momentum plane with the corresponding contours of trapezoidal shape around them discretized using $60$ mesh points. For the $A=10,11,12,15$ systems, the HO basis is used, as the protons and neutrons are very well bound.

\begin{figure*}[h!tb]
    \centering
    \includegraphics[width=0.325\textwidth]{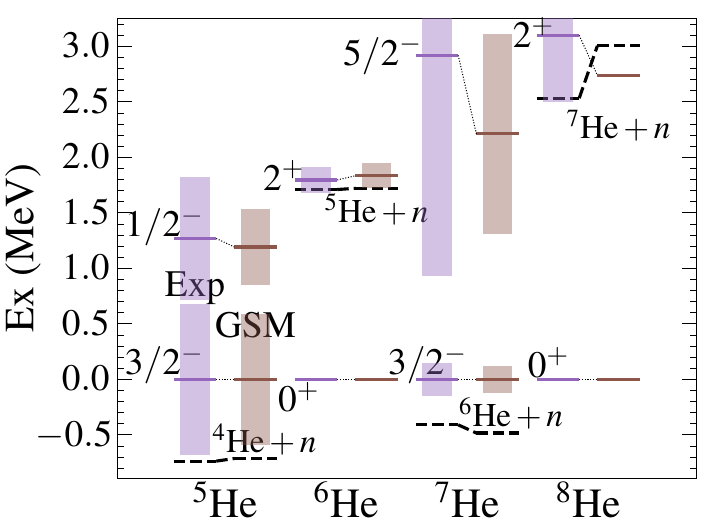}
    \includegraphics[width=0.325\textwidth]{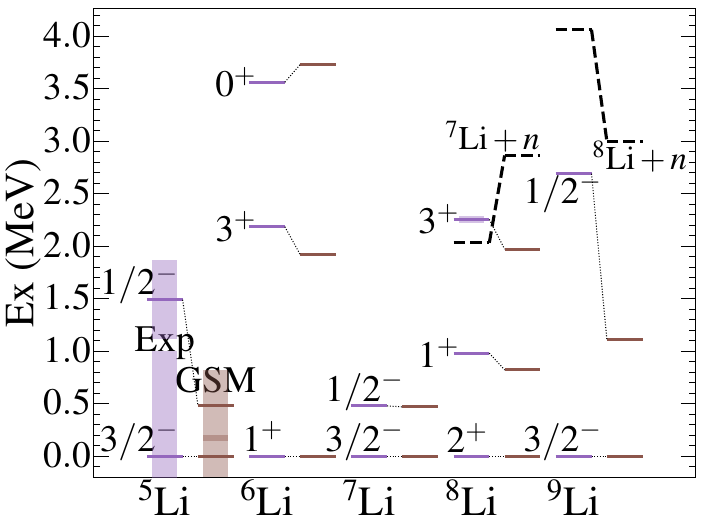}
    \includegraphics[width=0.325\textwidth]{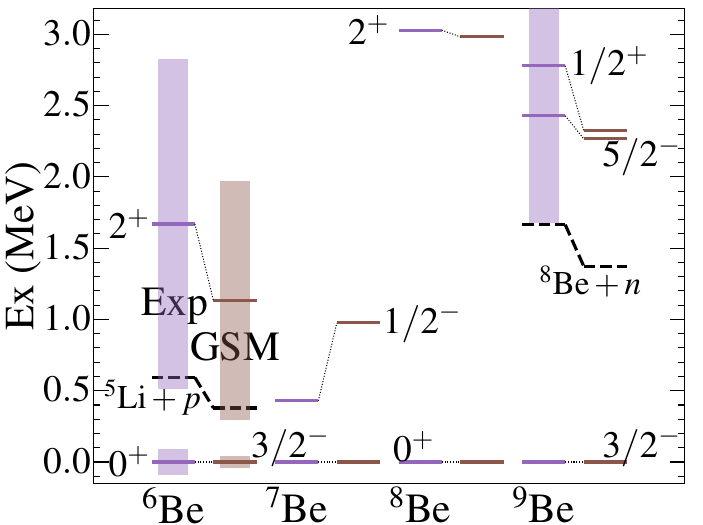}
    \caption{Spectra of isotopes of Helium (left), Lithium (middle) and Beryllium (right) calculated with GSM compared with the experimental data from Ref. \cite{nationalnucleardatacenter}. The widths of the excited states of $\nn{5}{He}$ and $\nn{5}{Li}$ are an order of magnitude smaller than the actual results ($0.1\Gamma$) to simplify the figure.}
    \label{fig:sec:nucleonsector:spectra}
\end{figure*}

The optimized strengths of the nucleon-nucleon interaction, which reproduce the binding energies and low-lying spectra of the selected nuclei, are shown in Table \ref{tab:sec:nucleonsector:tb_optimized}. Table \ref{tab:sec:nucleonic-part:optimizedenergies-widths} presents the calculated and experimental energies and widths of the optimized states. For the helium, lithium, and beryllium isotopes, a similar analysis was performed in Ref. \cite{jaganathenPhys.Rev.C2017} using the same Hamiltonian parameters for the three chains. However, since our goal is to have the best possible nucleon sector for constructing the hypernuclei, an independent Hamiltonian parametrization for each chain is adopted. 

For the helium and beryllium isotopes, the ground state energies deviate from experiment by less than $500\si{keV}$, with the largest error occurring for the heavier systems. The excited state energies are generally well reproduced, with neutron widths comparable to the experimental data. However, for the excited $1/2^-$ state of $\nn{9}{Be}$, although it is located more than $1\si{MeV}$ above the neutron threshold, its width is very small. This is a consequence of the limited size of the model space and the limited number of particle-hole excitations to the continuum due to computational limitations. For the lithium isotopes, the ground state energies for the first three lightest isotopes are in good agreement with the experiment, whereas the heavier isotopes differ by more than $500\si{keV}$. Figure \ref{fig:sec:nucleonsector:spectra} presents a comparison between the calculated excited states of the isotopes and the experimental data from Ref. \cite{nationalnucleardatacenter}. 

For the optimization of the mirror systems $\nn{11}{C}$ and $\nn{11}{B}$, in addition to the optimized spectra of these systems, the binding energies and selected states of the nuclei $\nn{10}{B}$ and $\nn{12}{C}$ are included to ensure a good nucleon separation energies with the optimized Hamiltonian. The ground state energies of the four nuclei are in excellent agreement with the experiment. This shows that the HO basis is an appropriate choice to describe the low-lying states of these well-bound systems. This is further confirmed by the accurate description of $\nn{15}{O}$ spectrum. The only state not described by our Hamiltonian is the $5/2^+$ state, with an experimental energy of $E_\mathrm{exp} = -78.419 \si{MeV}$, as it requires the inclusion of the $\spl{}{d}{5}$ pole in the model space.

\subsection{Determination of the $\Lambda$-core and $\Lambda N$ interaction strengths}  \label{sec:n-lambda_interaction} 

Selected $p$-shell hypernuclei (up to $\nhl{16}{O}$) are described within the GSM framework with the valence $\Lambda$-hyperon in the $\ell=0$ partial wave within the Berggren ensemble, assuming the $\Lambda N$ interaction is the same for all the studied systems. For the $\Lambda$-core Woods-Saxon potential, the reduced radius and diffusivity are fixed to $R_0 = 1.25A_\mathrm{core}^{1/3}\si{fm}$ and $a=0.65\si{fm}$, respectively. The depth is adjusted to $V_\mathrm{WS}^\Lambda = -25.147 \si{MeV}$ to reproduce the experimental binding energy of the hypernucleus $\nhl{5}{He}$ $E_\mathrm{Exp}(\nhl{5}{He}) = 3.170(16)\si{MeV}$ modeled as a $\ell=0$ $\Lambda$-hyperon above the $\nn{4}{He}$ core. For the Berggren basis diagonalizing potential, a bound $s$-pole is used with $30$ scattering states defined along the real axis of the complex momentum plane since all the studied states are bound with respect to $\Lambda$-emission.

To determine the parameters $\bar{v}^{\Lambda N}$ of each part of the $\Lambda N$ interaction, we follow the procedure used in the standard shell model \cite{millenerTopicsinStrangenessNuclearPhysics2007,galRev.Mod.Phys.2016}: first, the central and spin-orbit parts are adjusted to reproduce the first and second doublets of states in $\nhl{7}{L}$; second, the spin-orbit interaction is modified to reproduce the $5/2^+-3/2^+$ splitting in $\nhl{9}{Be}$; and finally, the tensor strength is adjusted to reproduce the ground state doublet splitting of $\nhl{16}{O}$. This process is repeated until a minimum is found. All calculations are performed with the $\nn{4}{He}$ inert core. 

The converged set of parameters is presented in Table \ref{tab:sec:lambdasector:tb_optimized}. The central part exhibits a dependence of the form $v_\alpha^{\Lambda N} = 0.81(-1)^{S+T}$ to reproduce the attractive and repulsive character of the EVEN and ODD parts, as observed in Ref. \cite{hiyamaPhys.Rev.C2006}. With the inclusion of the tensor force, the ratio $\eta_\mathrm{ALS}$ of the symmetric and antisymmetric spin-orbit parts converges to a value closer to that of the ESC08c potential \cite{veselyNuclearPhysicsA2016}.

\begin{table}[h!tb]
    \setlength{\tabcolsep}{0.5em}
    \renewcommand{\arraystretch}{1.4} 
	\caption{Strengths of the $\Lambda N$ central (C), spin-orbit (SO) and tensor (T) parts that multiplies the interaction parameters from Ref. \cite{hiyamaPhys.Rev.C2009}.}
	\label{tab:sec:lambdasector:tb_optimized}
	\begin{tabular}{c|c||c|c}\hline\hline
		$\bar{v}^\mathrm{C}_\alpha[\Lambda N]$ & MeV & $\bar{v}^\mathrm{LS/T}_\alpha[\Lambda N]$ & MeV \\ \hline
		$\bar{v}_\text{to}$ & $0.81$ & $\bar{v}^\mathrm{SO}_\text{to}$ & $0.70$ \\
		$\bar{v}_\text{te}$ & $-0.81$ & $\eta_\text{ALS}$ & $-0.61$\\
		$\bar{v}_\text{so}$ & $0.81$ & $\bar{v}^\mathrm{T}_\text{to}$ & $-2.50$ \\
		$\bar{v}_\text{se}$ & $-0.81$ & $\bar{v}^\mathrm{T}_\text{te}$ & $-2.50$ \\\hline\hline		
	\end{tabular}
\end{table}

In the following sections we will present the analysis of the results for all the hypernuclei under study.

\subsection{Binding energies and spectra of hypernuclei} \label{sec:hypernuclei-spectra}

With the nucleon-nucleon and $\Lambda N$ interactions determined, we now apply the GSM to calculate the binding energies and excitation spectra of selected hypernuclei.

\begin{table}[h!tb]
	\centering
    \renewcommand{\arraystretch}{1.5} 
	\caption{One-body WS strengths (in MeV) readjusted for the hypernuclei chain. Uncertainties are given in parentheses.}
	\label{tab:sec:lambdasector:ob_optimized}
	\begin{tabular}{c|c|c|c|@{\hspace{0.5em}}c@{\hspace{0.5em}}|@{\hspace{0.5em}}c@{\hspace{0.5em}}} \hline \hline
		$V_\text{WS}^\tau$ & $\nhl{\mathrm{A}}{He}$-chain & $\nhl{\mathrm{A}}{Li}$-chain & $\nhl{\mathrm{A}}{Be}$-chain & $\nhl{12}{C}/\nhl{12}{B}$ & $\nhl{16}{O}$ \\ \hline
		$V_\text{WS}^p$ & $-$           & $50.71(2)$ & $53.88(3)$ & $54.71(3)$ & $52.56$ \\
		$V_\text{WS}^n$ & $52.34(58)$ & $59.22(3)$ & $50.86(3)$ & $55.01(3)$ & $52.86$  \\\hline
		$V_\text{so}^p$ & $-$           & $1.77$      & $2.06$      & $5.38$ & $5.35$ \\
		$V_\text{so}^n$ & $5.56$       & $3.73$      & $4.35$      & $4.82$ & $4.79$ \\ \hline \hline
	\end{tabular}
\end{table}

The inclusion of the $\Lambda$-hyperon in the model space requires a modification of the nucleonic mean-field, as the model space is no longer the same \cite{millenerTopicsinStrangenessNuclearPhysics2007}. The Woods-Saxon strengths for the neutrons and protons from Table \ref{tab:sec:nucleonsector:ob_optimized} are readjusted to the values in Table \ref{tab:sec:lambdasector:ob_optimized} to reproduce the experimental binding energies of the selected hypernuclei, while the $NN$ interaction remains the same as in Table \ref{tab:sec:nucleonsector:tb_optimized}. For systems with similar numbers of valence protons and neutrons that are well bound respect to nucleon-emission thresholds, such as those with mass 12 and 16, the WS depths for protons and neutrons are very similar and close to the ``universal" value \cite{bohr1969}. The uncertainties in the new WS depths arise from the large experimental uncertainties for some hypernuclei. The isospin dependence given in eq. (\ref{eq:ws_isospin_dependence}) is applied with the same $\chi$ factors, and the isospin $I=(N-Z)/2$ corresponds to that of the nucleonic subsystem.  

\begin{table}[h!tb]
    \centering
    \renewcommand{\arraystretch}{1.32} 
    \caption{Calculated and experimental \cite{HypernuclearDataBase} energies (in MeV) with respect to the $\nn{4}{He}$ core and widths (in MeV) of all the calculated states of the selected hypernuclei. Uncertainties are given in parentheses.}
    \begin{tabular}{cc|@{\hspace{0.5em}}c@{\hspace{0.5em}} @{\hspace{0.5em}}c@{\hspace{0.5em}}|@{\hspace{0.3em}}c@{\hspace{0.3em}}}
        \hline\hline
        Hyper & State & $E_\mathrm{GSM}$ & $E_\mathrm{Exp}$ & $\Gamma_\mathrm{GSM}$  \\\hline
		$\nhl{6}{He}$ & $1^-$ & $-3.390(84)$ & $-3.444(100)$ & $-$ \\
		$\nhl{6}{He}$ & $2^-$ & $-3.238(76)$ & $-$ & $-$ \\
		$\nhl{6}{He}$ & $1^-$ & $-1.612(40)$ & $-$ & $2.359$ \\
		$\nhl{6}{He}$ & $0^-$ & $-1.583(43)$ & $-$ & $2.383$ \\
		$\nhl{7}{He}$ & $1/2^+$ & $-6.063(313)$ & $-6.034(333)$ & $-$   \\
		$\nhl{7}{He}$ & $3/2^+$ & $-3.862(280)$ & $-4.134(402)$ & $-$   \\
		$\nhl{7}{He}$ & $5/2^+$ & $-3.814(282)$ & $-$ & $-$   \\
		$\nhl{8}{He}$ & $1^-$ & $-8.179(356)$ & $-7.726(700)$ & $-$  \\
		$\nhl{8}{He}$ & $2^-$ & $-8.106(357)$ & $-$ & $-$  \\
		$\nhl{8}{He}$ & $0^-$ & $-6.696(258)$ & $-$ & $-$  \\
		$\nhl{8}{He}$ & $1^-$ & $-6.613(333)$ & $-$ & $-$  \\
		$\nhl{9}{He}$ & $1/2^+$ & $-10.492(660)$ & $-$ & $-$   \\
		$\nhl{9}{He}$ & $3/2^+$ & $-7.242(616)$ & $-$ & $0.001$   \\
		$\nhl{9}{He}$ & $5/2^+$ & $-7.186(614)$ & $-$ & $0.001$   \\ \hline
		$\nhl{6}{Li}$ & $1^-_1$ & $-2.468(28)$ & $-2.620(484)$ & $0.232$  \\
		$\nhl{6}{Li}$ & $2^-$   & $-2.367(28)$ & $-$ & $0.342$  \\
		$\nhl{6}{Li}$ & $0^-$   & $-1.684(25)$ & $-$ & $1.068$  \\
		$\nhl{6}{Li}$ & $1^-_2$ & $-1.676(24)$ & $-$ & $1.111$  \\
		$\nhl{7}{Li}$ & $1/2^+$ & $-9.091(18)$ & $-9.295(60)$ & $-$ \\
		$\nhl{7}{Li}$ & $3/2^+$ & $-8.401(17)$ & $-8.603(60)$ & $-$  \\
		$\nhl{7}{Li}$ & $5/2^+$ & $-7.038(18)$ & $-7.245(60)$ & $-$ \\
		$\nhl{7}{Li}$ & $7/2^+$ & $-6.556(17)$ & $-6.774(60)$ & $-$  \\
		$\nhl{8}{Li}$ & $1^-_1$ & $-17.607(30)$ & $-17.739(44)$ & $-$  \\
		$\nhl{8}{Li}$ & $2^-$   & $-17.155(29)$ & $-$ & $-$ \\ 
		$\nhl{8}{Li}$ & $0^-$   & $-16.760(29)$ & $-$ & $-$  \\
		$\nhl{8}{Li}$ & $1^-_2$ & $-16.594(29)$ & $-$ & $-$ \\ 
		$\nhl{9}{Li}$ & $3/2^+_1$ & $-21.946(38)$ & $-21.355(71)$ & $-$ \\
		$\nhl{9}{Li}$ & $5/2^+_1$ & $-21.280(38)$ & $-20.785(139)$ & $-$ \\
		$\nhl{9}{Li}$ & $1/2^+$   & $-21.192(39)$ & $-19.862(111)$ & $-$ \\
		$\nhl{9}{Li}$ & $3/2^+_2$ & $-20.664(38)$ &  & $-$ \\
		$\nhl{9}{Li}$ & $5/2^+_2$ & $-20.035(38)$ & $-19.085(115)$ & $-$ \\
		$\nhl{10}{Li}$ & $1^-_1$ & $-26.388(48)$ & $-$ & $-$  \\
		$\nhl{10}{Li}$ & $2^-$   & $-25.634(47)$ & $-$ & $-$ \\ 
		$\nhl{10}{Li}$ & $0^-$   & $-25.067(48)$ & $-$ & $-$  \\
		$\nhl{10}{Li}$ & $1^-_2$ & $-24.519(47)$ & $-$ & $-$ \\
        \hline\hline
    \end{tabular}
    \label{tab:sec:hypernuclei-part:optimizedenergies-widths}
\end{table}
\begin{table}[h!tb]
    \centering
    \renewcommand{\arraystretch}{1.32} 
    \caption{Continuation of Table \ref{tab:sec:hypernuclei-part:optimizedenergies-widths}.}
    \begin{tabular}{cc|@{\hspace{0.5em}}c@{\hspace{0.5em}} @{\hspace{0.5em}}c@{\hspace{0.5em}}|@{\hspace{0.5em}}c@{\hspace{0.5em}}}
        \hline\hline
        Hyper & State & $E_\mathrm{GSM}$ & $E_\mathrm{Exp}$ & $\Gamma_\mathrm{GSM}$  \\\hline
        $\nhl{7}{Be}$ & $1/2^+$ & $-3.827(89)$ & $-3.788(89)$ & $-$  \\
		$\nhl{7}{Be}$ & $3/2^+$ & $-2.262(71)$ & $-$ & $-$  \\
		$\nhl{7}{Be}$ & $5/2^+$ & $-2.251(74)$ & $-$ & $-$  \\
		$\nhl{8}{Be}$ & $1^-_1$ & $-15.837(33)$ & $-16.116(59)$ & $-$  \\
		$\nhl{8}{Be}$ & $2^-$   & $-15.457(33)$ & $-$ & $-$   \\
		$\nhl{8}{Be}$ & $0^-$   & $-14.443(32)$ & $-$ & $-$   \\
		$\nhl{8}{Be}$ & $1^-_2$ & $-14.383(32)$ & $-$ & $-$   \\
		$\nhl{9}{Be}$ & $1/2^+$ & $-35.066(48)$ & $-34.818(72)$ & $-$  \\
		$\nhl{9}{Be}$ & $5/2^+$ & $-32.195(49)$ & $-31.793(72)$ & $-$  \\
		$\nhl{9}{Be}$ & $3/2^+$ & $-32.151(49)$ & $-31.751(72)$ & $-$  \\
		$\nhl{10}{Be}$ & $1^-_1$ & $-37.836(57)$ & $-38.061(247)$ & $-$  \\
		$\nhl{10}{Be}$ & $2^-_1$ & $-37.618(57)$ & $-$ & $-$   \\ 
		$\nhl{10}{Be}$ & $2^-_2$ & $-35.563(57)$ & $-35.281(275)$ & $-$   \\ 
		$\nhl{10}{Be}$ & $3^-$   & $-35.382(57)$ & $-$ & $-$   \\ 
		$\nhl{10}{Be}$ & $0^-$   & $-34.880(57)$ & $-$ & $<0.001$   \\ 
		$\nhl{10}{Be}$ & $1^-_2$ & $-34.860(57)$ & $-$ & $<0.001$   \\ \hline
        $\nhl{12}{B}$ & $1^-_1$ & $-58.848(78)$ & $-59.328(64)$ & $-$\\
        $\nhl{12}{B}$ & $2^-$   & $-58.655(78)$ & $-59.149(86)$ & $-$ \\
        $\nhl{12}{B}$ & $0^-$   & $-56.058(78)$ & $-56.395(89)$ & $-$ \\
        $\nhl{12}{B}$ & $1^-_2$ & $-55.954(77)$ & $-$ & $-$ \\
        $\nhl{12}{C}$ & $1^-_1$ & $-56.867(78)$ & $-56.480(179)$ & $-$ \\
        $\nhl{12}{C}$ & $2^-$   & $-56.672(78)$ & $-56.318(179)$ & $-$ \\
        $\nhl{12}{C}$ & $0^-$   & $-54.140(77)$ & $-53.647(179)$ & $-$ \\
        $\nhl{12}{C}$ & $1^-_2$ & $-54.021(77)$ & $-$ & $-$ \\\hline
        $\nhl{16}{O}$ & $0^-$   & $-96.660$ & $-96.660(89)$ & $-$ \\
        $\nhl{16}{O}$ & $1^-_1$ & $-96.634$ & $-96.634(89)$ & $-$ \\
        $\nhl{16}{O}$ & $1^+$   & $-91.579$ & $-$ & $-$ \\
        $\nhl{16}{O}$ & $0^+$   & $-91.169$ & $-$ & $-$ \\
        $\nhl{16}{O}$ & $1^-_2$ & $-90.120$ & $-90.099(89)$ & $-$ \\
        $\nhl{16}{O}$ & $2^-$   & $-89.712$ & $-$ & $-$ \\
        \hline\hline
    \end{tabular}
    \label{tab:sec:hypernuclei-part:optimizedenergies-widths-2}
\end{table}

A particular distinction is required when the $\Lambda$-hyperon is coupled to an unbound nucleus, as the glue-like behavior of the hyperon introduces deeper WS potentials for the nucleons. This is the case for the systems $\nhl{6,8}{He}$, $\nhl{6}{Li}$, and $\nhl{7}{Be}$, where the WS strengths are $V_\mathrm{WS}^n = 54.4(4)$ and $V_\mathrm{WS}^p = 55.8(2)$, instead of those in Table \ref{tab:sec:lambdasector:ob_optimized}. Regarding the spin-orbit interaction for the $\nhl{\mathrm{A}}{Li}$ isotopes, as well as for mass 12 and $\nhl{16}{O}$ from Table \ref{tab:sec:nucleonsector:ob_optimized}, the strengths $V_\mathrm{so}^{n/p}$ are modified to reproduce the excitation energy of the first excited doublets of $\nhl{7}{Li}$, $\nhl{12}{C}$ and $\nhl{16}{O}$.

\begin{figure*}[h!tb]
    \centering
    \includegraphics[width=0.325\textwidth]{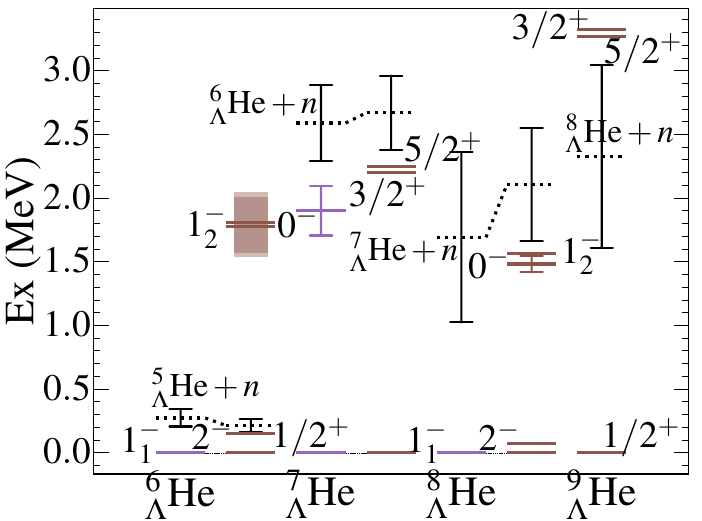}
    \includegraphics[width=0.325\textwidth]{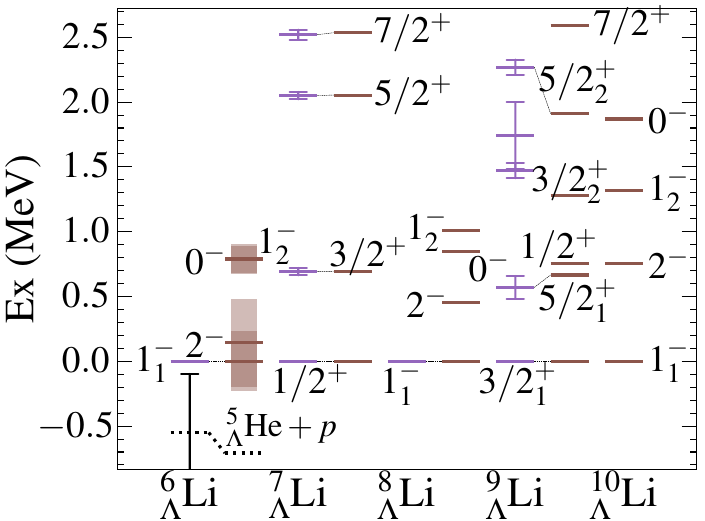}
    \includegraphics[width=0.325\textwidth]{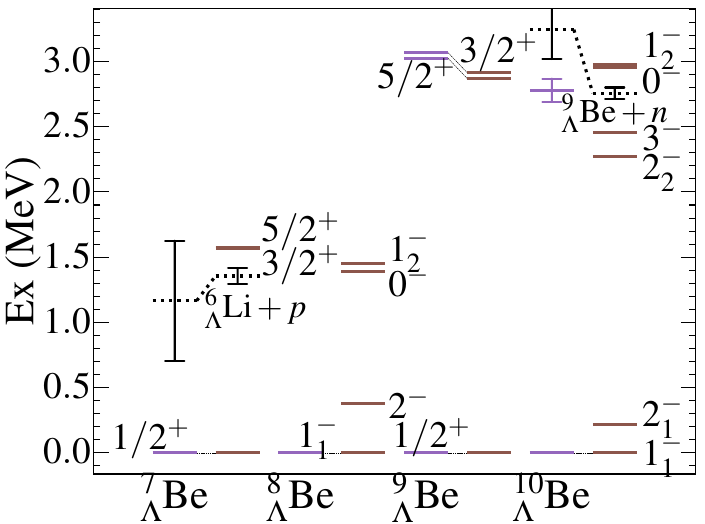}
    \caption{Spectra of hyper-isotopes of Helium (left), Lithium (center) and Beryllium (right) calculated with GSM are compared with the experimental data from Ref. \cite{nationalnucleardatacenter}. The widths of the represented states of $\nhl{6}{He}$ and $\nhl{6}{Li}$ in the figure are an order of magnitude smaller than the actual results ($0.1\Gamma$) to simplify the drawing.}
    \label{fig:sec:lambdasector:spectra}
\end{figure*}
 
The GSM energies, compared with the experiment, are presented in Tables \ref{tab:sec:hypernuclei-part:optimizedenergies-widths} and \ref{tab:sec:hypernuclei-part:optimizedenergies-widths-2}. We analyze the results for each hypernuclear set. The binding energies of the $\nhl{\mathrm{A}}{He}$ isotopes are in excellent agreement with experimental data. The determination of the neutron WS interaction for the $\nhl{9}{He}$ isotope is derived from that of $\nhl{7}{He}$, as it is the only $\nhl{\mathrm{A}}{He}$ isotope under consideration in this work in which the $\Lambda$-hyperon is coupled to a bound nucleonic subsystem with experimental data. Given the large uncertainty in this data, the predicted binding energy of the $\nhl{9}{He}$ also has a large uncertainty. 
This binding energy is in agreement with with a cluster model \cite{myoPhys.Rev.C2023} and a GSM calculation without the $\Lambda N - \Sigma N$ channel coupling \cite{liPhysicsLettersB2025}, but approximately $1\si{MeV}$ smaller than another from standard shell-model. 

The calculated ground state energies of the $\nhl{\mathrm{A}}{Li}$ isotopes reproduce the experimental data to within $300\si{keV}$, except for the heaviest isotope, $\nhl{9}{Li}$, for which our result is $\sim600\si{keV}$ more bound than the averaged value of Ref. \cite{HypernuclearDataBase}, and is closer to the emulsion experiment measurement \cite{PNIEWSKI1985685}. The $\nhl{6}{Li}$ isotope is proton-unbound, with a calculated ground-state width of $\Gamma(\nhl{6}{Li}) = 232\si{keV}$. The predicted binding energy of the $\nhl{10}{Li}$ isotope lies between the standard shell-model value \cite{galPhysicsLettersB2013} and a no-core shell-model calculation \cite{wirthPhysicsLettersB2018}. 

\begin{figure*}[h!tb]
    \centering
    \includegraphics[width=0.325\textwidth]{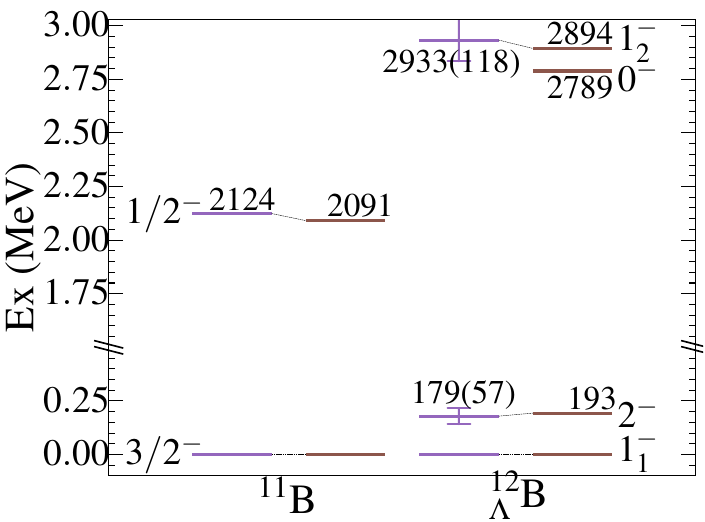}
    \includegraphics[width=0.325\textwidth]{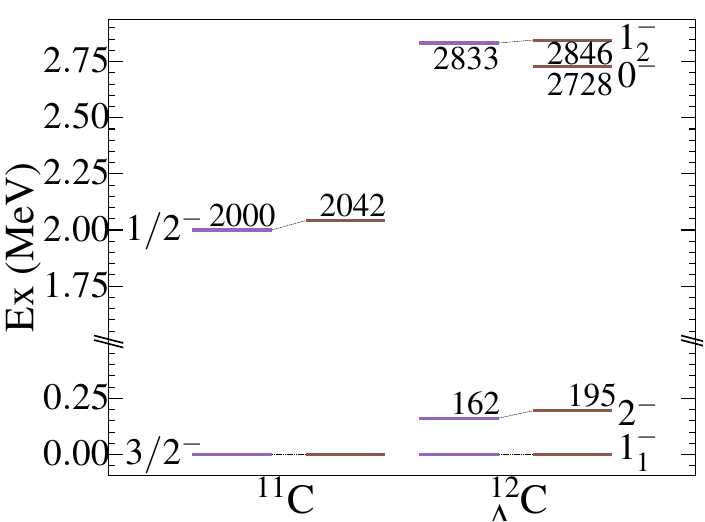}
    \includegraphics[width=0.325\textwidth]{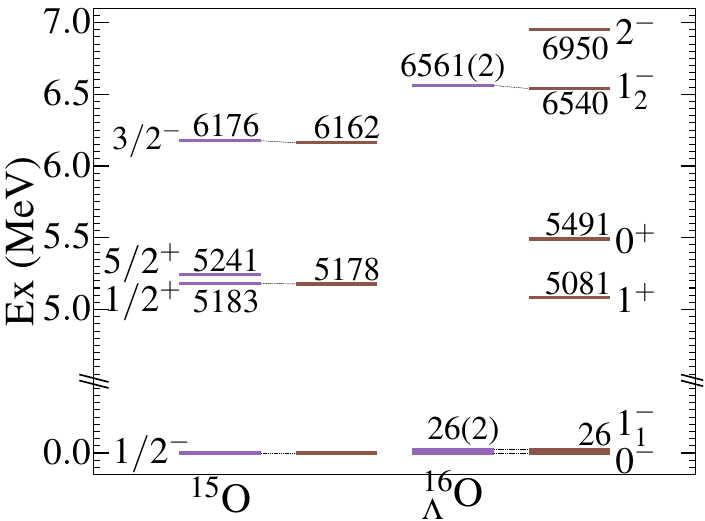}
    \caption{GSM spectra of $\nn{11}{B}$-$\nhl{12}{B}$ (left), $\nn{11}{C}$-$\nhl{12}{C}$ (center) and $\nn{15}{O}$-$\nhl{16}{O}$ (right) are compared with the experimental data \cite{nationalnucleardatacenter,HypernuclearDataBase}.}
    \label{fig:sec:12ALambda:spectra}
\end{figure*}

The results for the $\nhl{\mathrm{A}}{Be}$ isotopes, $\nhl{12}{C/B}$ and $\nhl{16}{O}$ are shown in Table \ref{tab:sec:hypernuclei-part:optimizedenergies-widths-2}. The calculated ground state energies for all the $\nhl{\mathrm{A}}{Be}$ isotopes differ by less than $300\si{keV}$ from experiment. In the particular case of $\nhl{7}{Be}$, where the $\Lambda$-hyperon is coupled to an unbound nucleonic subsystem, the same proton WS as in $\nhl{6}{Li}$ is used, and its calculated binding energy is in excellent agreement with the averaged value from Ref. \cite{HypernuclearDataBase}. 

For the hypernuclei $\nhl{12}{B}$ and $\nhl{12}{C}$, the same proton and neutron WS depths are used because the $\Lambda$-hyperon is coupled to the mirror systems $\nn{11}{B}$ and $\nn{11}{C}$. The calculated $S_\Lambda(\nhl{12}{B})=10.952\si{MeV}$ and $S_\Lambda(\nhl{12}{C})=11.697\si{MeV}$ using the averaged WS strengths are within $500\si{keV}$ of the average experimental data $S_\Lambda^\mathrm{exp}(\nhl{12}{B})=11.430(64)\si{MeV}$ and $S_\Lambda^\mathrm{exp}(\nhl{12}{C})=11.332(126)\si{MeV}$ \cite{HypernuclearDataBase}. Finally, the ground state of $\nhl{16}{O}$ is exactly reproduced with WS strengths smaller than those for the mirror systems, as noted in Table \ref{tab:sec:lambdasector:ob_optimized}.

\begin{table*}[h!tb]
    \centering
    \renewcommand{\arraystretch}{1.2} 
    \caption{Energy spacings (in keV) between states in selected hypernuclei. $\Delta E_C$ is the contribution of the nucleon sector level spacing, EVEN, ODD  are the contributions from each part of the central $\Lambda N$ interaction, SLS and ALS are the contributions from the spin-orbit parts, and T from the Tensor part. The last row for each hypernucleus indicates the contribution to the binding energy from the different interaction components. Experimental uncertainties are given in parentheses.}
    \begin{tabular}{cc|@{\hspace{0.5em}}c@{\hspace{0.5em}} ccccc|rr}
        \hline\hline
        & $J_u^\pi-J_l^\pi$ & $\Delta E_C$ & +EVEN & +ODD & +SLS & +ALS & +T & $\Delta E_\mathrm{GSM}$ & $\Delta E_\mathrm{exp}$ \\ \hline
        $\nhl{6}{He}$ 
        & $2^--1^-_1$   & $0$    & $276$ & $-99$  & $-52$ & $51$  & $-24$ & $152$  & \\
        & $1^-_2-1^-_1$ & $1526$ & $268$ & $-81$  & $3$   & $74$  & $-12$ & $1777$ & \\
        & $0^--1^-_2$ & $0$    & $57$  & $-20$  & $26$  & $-23$ & $-11$ & $30$   & \\
        & BE          & $-$    & $331$ & $-112$ & $-13$ & $51$  & $-15$ & $242$  & \\\hline
        $\nhl{9}{Li}$ 
        & $5/2^+_1-3/2^+_1$ & $0$    & $1020$ & $-366$ & $-98$  & $80$  & $17$  & $653$  & $570(120)$ \\
        & $1/2^+-3/2^+_1$   & $666$ & $-79$ & $-3$ & $24$ & $147$ & $-2$ & $753$\\
        & $3/2^+_2-3/2^+_1$ & $666$  & $809$  & $-251$ & $-9$   & $99$  & $-45$ & $1269$ & $1470(90)$ \\
        & $5/2^+_2-3/2^+_1$ & $1993$ & $-95$  & $-32$  & $-29$  & $57$  & $16$  & $1910$ & $2270(70)$ \\
        & BE              & $-$    & $2224$ & $-674$ & $52$   & $124$ & $12$  & $1739$ & \\\hline
        $\nhl{9}{Be}$
        & $5/2^+-1/2^+$ & $2961$ & $-57$ & $15$ & $-114$ & $102$  & $-11$ & $2895$ & $3024(3)$  \\
        & $3/2^+-5/2^+$ & $0$    & $-30$ & $13$ & $283$  & $-248$ & $27$  & $45$   & $43(3)$  \\ 
        & BE            & $-$    & $1881$ & $-506$ & $43$ & $40$ & $4$ & $1463$ &  \\\hline
        $\nhl{12}{C}$ 
        & $2^--1^-_1$  & $0$    & $487$  & $-185$  & $-174$ & $156$ & $-89$ & $195$  & $162(1)$ \\
        & $1^-_2-1^-_1$ & $2305$ & $409$  & $-167$  & $13$   & $241$ & $38$  & $2839$ & $2833(2)$ \\
        & $1^-_2-0^-$ & $0$    & $-252$ & $79$    & $-136$ & $115$ & $312$ & $118$  &  \\
        & BE          & $-$    & $3614$ & $-1033$ & $213$  & $368$ & $-41$ & $3122$ & $-$\\\hline
        $\nhl{16}{O}$ 
        & $1^-_1-0^-$ & $0$    & $-569$ & $214$   & $-155$ & $133$ & $403$  & $26$   & $26(2)$ \\
        & $1^-_2-0^-$ & $6415$ & $-803$ & $287$   & $123$  & $135$ & $377$  & $6540$ & $6561(2)$ \\
        & $2^--1^-_2$ & $0$    & $801$  & $-270$  & $-151$ & $130$ & $-102$ & $408$  &  \\
        & BE          & $-$    & $4612$ & $-1213$ & $34$   & $222$ & $249$  & $3904$ & \\\hline\hline
    \end{tabular}
    \label{tab:yn-interaction-dependence}
\end{table*}

Figure \ref{fig:sec:lambdasector:spectra} (left) shows the spectra of the helium hyperisotopes compared to experimental data from Ref. \cite{HypernuclearDataBase}, where available. The calculated excited doublet of $\nhl{6}{He}$ is neutron-unbound, with smaller widths than those calculated with a different YN interaction \cite{myoPhys.Rev.C2023,liPhysicsLettersB2025}. The order of the states in the excited doublet matches that of the cluster model calculation \cite{myoPhys.Rev.C2023}, with a smaller separation energy due to the inclusion of the tensor force, as shown in Table \ref{tab:yn-interaction-dependence}. The experimentally observed excited state in $\nhl{7}{He}$ corresponds to the calculated $3/2^+-5/2^+$ doublet below the neutron threshold. The order of the states in the excited doublet agrees with different calculations \cite{myoPhys.Rev.C2023,liPhysicsLettersB2025}, but we predict a significantly smaller separation energy. The inclusion of the tensor force reduces the energy splitting from $80\si{keV}$ to $50\si{keV}$, compared to the cluster model calculation with the same $\Lambda N$ central and spin-orbit interaction. 

For $\nhl{8}{He}$, the calculated excited doublet is bound with respect to neutron emission, and the order of the states in this doublet is flipped with respect to the excited unbound doublet of $\nhl{6}{He}$. The tensor force effect in $\nhl{6}{He}$ is washed out by the continuum as it is an order of magnitude smaller than that of $\nhl{8}{He}$. The same situation is observed between the $3/2^+-5/2^+$ doublets in $\nhl{7}{He}$ and $\nhl{9}{He}$, where the tensor force energy splitting is an order of magnitude larger when the doublet is bound with respect to neutron emission. The excited doublet $3/2^+-5/2^+$ in $\nhl{9}{He}$ is predicted to be neutron-unbound, in agreement with different models \cite{myoPhys.Rev.C2023,liPhysicsLettersB2025}.

Figure \ref{fig:sec:lambdasector:spectra} shows the lithium hyperisotopes (center) compared to experimental data from Ref. \cite{HypernuclearDataBase}. The complete spectrum of $\nhl{6}{Li}$ is proton-unbound, with widths of the excited doublet half as large as those in $\nhl{6}{He}$. Furthermore, the tensor force between proton-$\Lambda$ generates an splitting four times larger than that for neutron-$\Lambda$ in $\nhl{6}{He}$, causing the flipped order of the excited states. 

The spectrum of $\nhl{7}{Li}$ was included in the fit of the $\Lambda N$ interaction, and a good reproduction of the four experimentally observed states is obtained. The other odd-A hypernucleus in this chain is $\nhl{9}{Li}$, where the splitting generated by the $\Lambda N$ interaction is larger than that observed in $\nhl{7}{Li}$. Furthermore, the splitting is large enough to mix the $5/2^+_1$ and $1/2^+$ states, which belong to different $\Lambda$-coupling configurations, at an energy close to the experimentally observed state. In Table \ref{tab:yn-interaction-dependence}, one can see how the splittings in this hypernucleus are generated: the position of the $5/2^+_1$ is determined by the $\Lambda N$ interaction, while the excitation energy of the $1/2^+$ is mainly determined by the nucleon sector. Indeed, this almost complete mixing between two doublets arises from two causes: a large $\Lambda N$ splitting and a small separation energy of the states in the nucleonic subsystem. Finally, a comparison of the excited doublets of $\nhl{6}{Li}$ and $\nhl{8}{Li}$ reveals that the nucleon continuum affects the $\Lambda N$ interaction energy splitting through the tensor force, as the tensor-force-induced splitting is five times larger for the bound $1^-_2-0^-$ doublet of $\nhl{8}{He}$ than for the unbound $1^-_2-0^-$ doublet of $\nhl{6}{He}$.

Figure \ref{fig:sec:lambdasector:spectra} (right) shows he spectra of the $\nhl{\mathrm{A}}{Be}$ isotopes. The excited doublet of $\nhl{7}{Be}$ is unbound with respect to proton emission, with its states separated by only $11\si{keV}$. This splitting arises from a combination of the central and tensor parts of the $\Lambda N$ interaction, because the symmetric and antisymmetric spin-orbit components are suppressed, as they contribute with equal magnitude but opposite sign. This is the opposite of what is observed for $\nhl{9}{Be}$ in Table \ref{tab:yn-interaction-dependence}, where the central and tensor parts cancel out, and the spin-orbit part is responsible for the energy splitting. Finally, as observed for the excited $1^-_2-0^-$ doublet in the $\nhl{6,8}{He}$ and $\nhl{6,8}{Li}$ systems, between the bound $\nhl{8}{Be}$ and the unbound $\nhl{10}{Be}$, the tensor-force-induced energy splitting is twice as large for the former as for the latter.

We calculate the mirror nuclei $\nn{11}{B}$ and $\nn{11}{C}$ using the Hamiltonian parameters of Tables \ref{tab:sec:nucleonsector:ob_optimized} and \ref{tab:sec:nucleonsector:tb_optimized}. The calculated excitation energy of the $1/2^-$ state in each nucleus, compared with the experiment, is shown in Figure \ref{fig:sec:12ALambda:spectra} (left and center). The GSM energy of the $1/2^-$ state in $\nn{11}{B}$ is larger than that in $\nn{11}{C}$, in agreement with experiment, but their separation is $49\si{keV}$ instead of $124\si{keV}$. This affects the calculated excited doublet of the hypernuclei $\nhl{12}{B}$ and $\nhl{12}{C}$. Then, the $V_\mathrm{so}^{p/n}$ should be increased by $15\%$, as shown in Table \ref{tab:sec:lambdasector:ob_optimized}, to place the $1^-_2$ state of $\nhl{12}{C}$ at the experimental energy, while the corresponding state of $\nhl{12}{B}$ is located $48\si{keV}$ above. Concerning the energy splitting due to the $\Lambda N$ interaction, our parametrization gives a splitting for the $\nhl{12}{C}$ ground state doublet that is $33\si{keV}$ larger than the experiment. From Table \ref{tab:yn-interaction-dependence} it can be seen that a $10\%$ smaller central $\Lambda N$ interaction would correct this difference, as noted in Ref. \cite{millenerNuclearPhysicsA2013}. 

With respect to the ground state doublet in $\nhl{12}{B}$, the calculated difference with the $\nhl{12}{C}$ hypernucleus is negligible, and we therefore predict that the ground-state energy splitting in $\nhl{12}{B}$ should be similar to that measured in $\nhl{12}{C}$. Regarding the excited doublet $1^-_2-0^-$, our model predicts the $0^-$ state as the first member of the doublet, in agreement with observations for the previous hypernuclear chains.

Figure \ref{fig:sec:12ALambda:spectra} (right) shows the calculated and experimental spectra of $\nn{15}{O}$ and $\nhl{16}{O}$. The $\Lambda N$ tensor interaction was adjusted to reproduce the energy splitting of the ground state doublet. As shown in Table \ref{tab:yn-interaction-dependence}, out stronger tensor force is used to compensate for the large splitting generated in this doublet by the central $\Lambda N$, in contrast to standard shell-model work \cite{millenerNuclearPhysicsA2008}. The tensor force required to describe the ground-state doublet splitting in $\nhl{16}{O}$ suggests that tensor contributions will remain essential for heavier hypernuclei, where the $\Lambda N$ interaction may exhibit even more complex spin-dependent behavior. 
As for the position of the excited $1^-_2$ state in $\nhl{16}{O}$, a $V_\mathrm{so}^{p/n}$ similar to that for the mirrors $\nhl{12}{B/C}$ is used (see Table \ref{tab:sec:lambdasector:ob_optimized}); the $2^-$ state located at $408\si{keV}$ due to the central $\Lambda N$ value. This result does not support the $2^-$ assignment to the E930 transition at the excitation energy $\mathrm{E}_\mathrm{exp} = 6758\pm4\pm4\si{keV}$ done by Ref. \cite{tamuraNuclearPhysicsA2008,millenerNuclearPhysicsA2008}.

The delicate balance between the $\Lambda N$ interaction and the nucleonic environment is a general feature of hypernuclei and will be crucial for understanding heavier systems. In particular, the competition between bound states and continuum coupling will shape both the spectrum and the stability of multi-strangeness hypernuclei.

\subsection{Densities and radii} \label{sec:hypernuclei-densities}
Beyond energies and spectra, the GSM also provides insights into the spatial structure of hypernuclei through single-particle densities and root-mean-square radii. Previous studies \cite{hiyamaPhys.Rev.C2009,liPhysicsLettersB2025,knollPhysicsLettersB2026} have shown that in the hypernuclei, since the $\Lambda$-hyperon is not subject to the Pauli exclusion principle with respect to nucleons, it can occupy the $s$-wave and be located inside the nucleonic cloud. In the $p$-shell hypernuclei, the $\Lambda$-hyperon is expected to have a different density distribution, because the low lying spectra of hypernuclei are primarily described as an $\ell=0$ $\Lambda$-hyperon coupled to $p$-wave nucleons. As discussed before, the $\Lambda$-hyperon is effectively treated in a $s$-wave in our model, but its wave function can be form from either the HO basis or the Berggren basis. 
\begin{figure}[h!tb]
    \centering
    \includegraphics[width=0.45\textwidth]{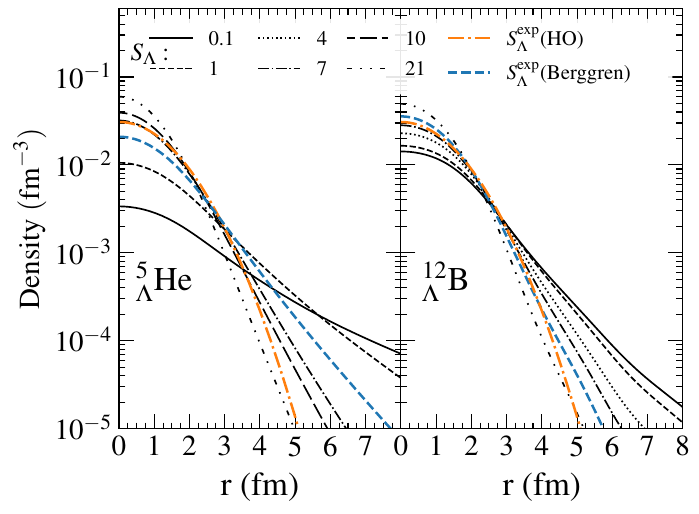}
    \caption{$\Lambda$-densities of two hypernuclei calculated with the Berggren basis for selected $S_\Lambda$ (black lines). In blue dashed lines the $\Lambda$-densities for the experimental $S_\Lambda^\mathrm{exp}(\nhl{5}{He})=3.102(30)\si{MeV}$ and $S_\Lambda^\mathrm{exp}(\nhl{12}{B})=11.430(64)\si{MeV}$ calculated with the Berggren basis, while in orange dash-dot lines calculated with the HO basis.}
    \label{fig:sec:observables:berggren-vs-HO}
\end{figure}

Figure \ref{fig:sec:observables:berggren-vs-HO} shows two selected examples of $\Lambda$-hyperon densities in hypernuclei. On the left, we show the simplest case, where the $\Lambda$-hyperon is the only valence particle. One can see how the $\Lambda$-density, calculated with the Berggren basis, evolves from a weekly bound system ($S_\Lambda = 0.1\si{MeV}$), exhibiting the characteristic behavior of a halo system. For the experimental value $S_\Lambda^\mathrm{exp}(\nhl{5}{He})=3.102(30)\si{MeV}$, the Berggren-basis density clearly differs from the HO result, not only at large distances but also inside the nucleus. Furthermore, there are not densities calculated with the Berggren basis that agrees with the HO one for all radius.
\begin{figure}[h!tb]
    \centering
    \includegraphics[width=0.45\textwidth]{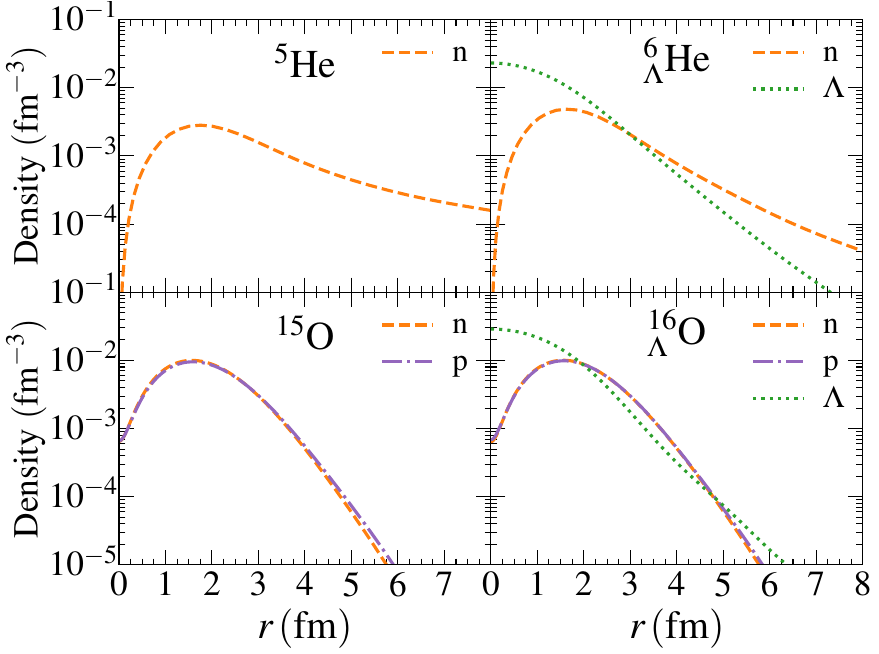}
    \caption{Single particle densities of selected hypernuclei compared with the ones of the nuclear subsystem.}
    \label{fig:sec:observables:particledensities}
\end{figure}

Now, we consider the system with additional valence neutrons and protons to describe the $\nhl{12}{B}$ hypernucleus (Figure \ref{fig:sec:observables:berggren-vs-HO}, right). It is clear that the $\Lambda N$ interaction leads to a shrinking of the $\Lambda$-hyperon density, with significant effects at small $S_\Lambda$, preventing halo-like behavior of the hyperon. Furthermore, at the experimental separation energy $S_\Lambda^\mathrm{exp}(\nhl{12}{B}) = 11.430(64)\si{MeV}$, the differences between the Berggren and HO densities are smaller than in the $\nhl{5}{He}$ case, but still none of the densities calculated with the Berggren basis reproduce exactly the HO one.

\begin{table*}[h!tb]
	\centering
    \renewcommand{\arraystretch}{1.2} 
	\caption{Calculated rms radii (in fm) for each baryon ($r_n$, $r_p$ and $r_\Lambda$) for the hypernuclei of the helium-chain ($\nn{\mathrm{A}}{He}$), lithium-chain ($\nn{\mathrm{A}}{Li}$), beryllium-chain ($\nn{\mathrm{A}}{Be}$) and mass equal 12 and 16 (A=12,16). For the radius of $^4$He-core,it was taken $r_\alpha=1.65\si{fm}$.}
	\label{tab:sec:gsm:ssec:25.12.15-10.00:rms_values}
	\begin{tabular}{c|@{\hspace{0.5em}}c@{\hspace{0.5em}}c||c|@{\hspace{0.5em}}c@{\hspace{0.5em}}cc||c|@{\hspace{0.5em}}c@{\hspace{0.5em}}cc||c|@{\hspace{0.5em}}c@{\hspace{0.5em}}cc}
        \hline\hline
		$\nn{\mathrm{A}}{He}$ & $r_n$ & $r_\Lambda$ & $\nn{\mathrm{A}}{Li}$ & $r_n$ & $r_p$ & $r_\Lambda$ & $\nn{\mathrm{A}}{Be}$ & $r_n$ & $r_p$ & $r_\Lambda$ & A=12,16 & $r_n$ & $r_p$ & $r_\Lambda$ \\\hline
		$\nhl{5}{He}$ & $-$ &  $2.352$ &  &  &  & &  &  &  &  &  &  &  & \\\hline
		$\nn{5}{He}$  & $2.884-i0.319$ & $-$ & 
        $\nn{5}{Li}$ & $-$ & $4.284+i0.561$ & $-$ &  &  &  &   &  &  &  & \\
		$\nhl{6}{He}$ & $3.040$  & $2.480$  & 
        $\nhl{6}{Li}$ & $-$ & $2.789-i0.020$ & $2.552$  &  &  &  &  &  &  &  & \\ \hline
		$\nn{6}{He}$ & $2.543$ &  $-$ & 
        $\nn{6}{Li}$ & $2.148$ & $2.233$ & $-$ & 
        $\nn{6}{Be}$ & $-$ & $3.270+i0.615$ & $-$    &  &  &  & \\
		$\nhl{7}{He}$ & $2.501$ & $2.265$  & 
        $\nhl{7}{Li}$ & $2.124$ & $2.207$ & $2.358$ & 
        $\nhl{7}{Be}$ & $-$ & $2.662$ & $2.650$  &  &  &  & \\\hline
		$\nn{7}{He}$ & $2.837-i0.177$ & $-$ & 
        $\nn{7}{Li}$ & $2.179$ & $2.118$ & $-$ & 
        $\nn{7}{Be}$ & $2.160$ & $2.340$ & $-$  &
        $\nn{11}{B}$ & $2.233$ & $2.166$ & \\
		$\nhl{8}{He}$ & $2.669$  & $2.238$ & 
        $\nhl{8}{Li}$ & $2.167$ & $2.094$ & $2.270$ & 
        $\nhl{8}{Be}$ & $2.261$ & $2.455$ & $2.309$  &  
        $\nhl{12}{B}$ & $2.219$ & $2.154$ & $2.113$ \\\hline
		$\nn{8}{He}$ & $2.744$  & $-$ & 
        $\nn{8}{Li}$ & $2.445$ & $2.157$ & $-$  & 
        $\nn{8}{Be}$ & $2.450$ & $2.480$ & $-$  & 
        $\nn{11}{C}$ & $2.156$ & $2.254$ & \\
		$\nhl{9}{He}$ & $2.695$ & $2.260$ & 
        $\nhl{9}{Li}$ & $2.392$ & $2.116$ & $2.245$ & 
        $\nhl{9}{Be}$ & $2.567$ & $2.609$ & $2.384$  & 
        $\nhl{12}{C}$ & $2.144$ & $2.237$ & $2.201$ \\\hline
		& &  & 
        $\nn{9}{Li}$ & $2.599$ & $2.171$ & $-$ & 
        $\nn{9}{Be}$ & $2.944$ & $2.616$ & $-$   &  
        $\nn{15}{O}$ & $2.282$ & $2.391$  & \\
		& &  & 
        $\nhl{10}{Li}$ & $2.538$ & $2.120$ & $2.269$ & 
        $\nhl{10}{Be}$ & $3.085$ & $2.742$ & $2.301$  &
        $\nhl{16}{O}$ & $2.269$ & $2.375$ & $2.282$ \\
        \hline\hline
	\end{tabular}
\end{table*}

Figure \ref{fig:sec:observables:particledensities} shows the single-particle densities of the valence baryons for two selected hypernuclei and their respective nucleonic subsystems. The neutron density in $\nn{5}{He}$ exhibits the characteristic behavior of a resonance, as described within the Berggren basis, with a maximum of $3\sn{-3}\si{fm^{-3}}$ at $r=1.8\si{fm}$; at $r=10\si{fm}$, the neutron density remains above $10^{-4}\si{fm^{-3}}$. 
When the $\ell=0$ $\Lambda$-hyperon is coupled to the $\nn{5}{He}$ nuclei, the resulting hypernucleus is bound with respect to neutron emission, with $S_n=220\si{keV}$, and the neutron density exhibits halo behavior. The maximum density now occurs at $r=1.6\si{fm}$ with a value of $5\sn{-3}\si{fm^{-3}}$. 
The $\Lambda$-density in $\nhl{6}{He}$ has an $s$-wave shape with a strong presence inside the hypernucleus surface $r_s=1.25\mathrm{A}^{1/3}\si{fm}=2.27\si{fm}$, corresponding to a well bound $\Lambda$-hyperon ($S_\Lambda=4.102\si{MeV}$). However, it also extends far beyond the nuclear surface, with a density exceeding $10^{-4}\si{fm^{-3}}$ even at $r=5\si{fm}$.

When the $\Lambda$-hyperon couples to a bound nucleus, such as $\nn{15}{O}$, a different situation arises. The proton and neutron densities of the $\nn{15}{O}$ nucleus show the feature of well-bound nucleons, which were described with the HO basis. When the $\Lambda$-hyperon is added using the $s$-wave Berggren basis, its density behaves similarly to that observed in $\nhl{6}{He}$. However, as observed in Figure \ref{fig:sec:observables:berggren-vs-HO}, it shrinks due to the presence of many valence nucleons, such that at the nuclear surface $r_s=1.25\mathrm{A}^{1/3}\si{fm}=3.15\si{fm}$ the neutron and proton densities are nine times larger than that of the $\Lambda$-hyperon. These insights into the radial dependence of baryons provide a baseline for exploring the spatial properties of heavier and double-strangeness hypernuclei, where the interplay between multiple strange baryons and the nucleonic core will be even more complex.

Finally, Table \ref{tab:sec:gsm:ssec:25.12.15-10.00:rms_values} shows the calculated root-mean-square (rms) radii of the studied hypernuclei and nuclei. The rms radii for resonant states are complex, and the imaginary part can be understood as its uncertainty \cite{berggrenPhysicsLettersB1996}.

When the $\Lambda$-hyperon is coupled to the lightest unbound nuclei ($\nn{5}{He}$, $\nn{5}{Li}$ and $\nn{6}{Be}$), the resulting hypernucleus is stable, with a smaller nucleon rms radius, while the $r_\Lambda$ values of these systems are the largest of our calculations and can be labeled as $\Lambda$-halos. For the $\nhl{\mathrm{A}}{He}$ isotopes, the relation $r_n > r_\Lambda$ is observed, as in other studies \cite{hiyamaPhys.Rev.C2009,liPhysicsLettersB2025}, while for the $\nhl{7,8}{Li}$ isotopes, $r_\Lambda$ is the largest of the three rms radii. Furthermore, in the $\nhl{\mathrm{A}}{Li}$ hypernuclei, the valence $\Lambda$-hyperon has always a larger rms radius than the valence proton. For the heaviest studied $p$-shell hypernuclei, the $\Lambda$-hyperon has an smaller rms $r_\Lambda$ than that of the protons and neutrons, and the probability of having a $\Lambda$-halo in heavier systems seems remote.

\section{Summary and conclusions} \label{sec:conclusions}
In this work, we have presented a systematic application of the GSM to the study of light $p$-shell hypernuclei, focusing on the role of the $\Lambda N$ interaction and its components within a unified framework. By extending the GSM to include hyperons as valence particles, we have demonstrated its capability to describe both bound and unbound states, as well as resonant and continuum effects, which are critical for systems near or beyond particle emission thresholds. The use of the Berggren basis, which incorporates bound states, resonances, and scattering continua, has allowed us to treat all quantum states on an equal footing and to explore the unique structural features of hypernuclei, such as halo and skin configurations.

Our calculations have provided a consistent and accurate description of the binding energies, excitation spectra, and density distributions for a wide range of hypernuclei, from light $\nhl{\mathrm{A}}{He}$, $\nhl{\mathrm{A}}{Li}$ and $\nhl{\mathrm{A}}{Be}$ isotopes to heavier systems like $^{12}_{\Lambda}\mathrm{B}$, $^{12}_{\Lambda}\mathrm{C}$, and $^{16}_{\Lambda}\mathrm{O}$. A unique $\Lambda N$ interaction was used for all these systems, with the central, spin-orbit and tensor parts determined by the doublets energy splitting in $^{7}_{\Lambda}\mathrm{Li}$, $^{9}_{\Lambda}\mathrm{Be}$ and $^{16}_{\Lambda}\mathrm{O}$, respectively. The inclusion of the tensor force in the $\Lambda N$ interaction was found to have a significant impact on the excited state splittings, particularly in bound systems, while its effect was washed out by the continuum in unbound cases. This highlights the delicate interplay between the $\Lambda N$ interaction and the nucleonic environment, where the presence of the continuum can suppress or enhance certain components of the interaction.

Our analysis of baryon radial distributions in selected hypernuclei reveals that $\Lambda$-densities calculated with the Berggren basis differs from HO results, even for large $\Lambda$-hyperon separation energies. The Berggren basis is essential for understanding the effective behavior of the $\Lambda$-hyperon, including its long-range tails at large distances and its distribution within the nucleon cloud.

This study analyzed the known role of the $\Lambda$-hyperon as a ``glue-like'' particle, which stabilizes weakly bound or resonant systems and leads to a shrinkage of the hyperon density in the presence of multiple valence nucleons. This effect is particularly evident in the radial densities and root-mean-square radii of the studied hypernuclei. For instance, the $\Lambda$ density in $^{6}_{\Lambda}\mathrm{He}$ exhibits a long-range behavior characteristic of halo systems, while in heavier systems like $^{12}_{\Lambda}\mathrm{B}$, the $\Lambda$ density is more localized due to the $\Lambda N$ interaction and the presence of additional valence nucleons. 

The systematic application of the GSM to $p$-shell hypernuclei has revealed the critical role of both the $\Lambda N$ interaction shaping hypernuclear structure and the nucleonic mean-field for the stabilization of the hypernuclei. This highlights a delicate interplay between the $\Lambda N$ interaction and the nucleonic environment.
The consistency of our results across the $p$-shell hypernuclei suggests that the optimized $\Lambda N$ interaction parameters and its structural features can be applied in heavier systems as well. Indeed, it is planned to measure energy spectra of $^{40,48}_{\Lambda}$K using $^{40,48}$Ca target by $(e,e'K^+)$ reaction at JLab. The GSM formalism and the $\Lambda N$ interaction presented in this work,  will be used in the forthcoming paper to calculate energy spectra and binding energies in these hypernuclei.

\begin{acknowledgments}
This work has been partially supported by JSPS Grants-in-Aid for Transformative Research Areas (Quantum Matter Science in the Universe Opened Up by Precise Numerical Calculations), JP25A203 and JP25H01267, by JST ERATO Grant Number JPMJER2304, and by the grants of the National Natural Science Foundation of China Nos. 12347106 and 12575124. We gratefully acknowledge support from the CNRS/IN2P3 Computing Center (Lyon, France) and the CRIANN (Normandy, France) for providing computing and data-processing resources needed for this work. 

\end{acknowledgments}

\bibliography{references}

\end{document}